\documentclass[10pt,final,journal,twocolumn]{IEEEtran}
\ifCLASSINFOpdf

\else
\usepackage{algorithm}
\usepackage{booktabs}

\fi
\usepackage{cite}
\usepackage[cmex10]{amsmath}
\usepackage{amsthm}
\usepackage{stfloats}
\usepackage{multicol,multienum}
\usepackage{multirow}

\usepackage{amssymb}
\usepackage{url}
\usepackage{amsmath}
\usepackage{xcolor}
\usepackage[dvips]{graphicx}
\usepackage{subfigure}
\usepackage{caption}
\usepackage{amsmath}
\usepackage{amsfonts,amssymb}
\usepackage{bbm}
\usepackage[T1]{fontenc}
\usepackage{algorithm}
\usepackage{algpseudocode}
\usepackage{ulem}
\usepackage[linesnumbered, ruled]{}
\makeatletter
\renewcommand{\maketag@@@}[1]{\hbox{\m@th\normalsize\normalfont#1}}%
\makeatother

\begin{document}
% correct bad hyphenation here
\hyphenation{op-tical net-works semi-conduc-tor}
%\begin{document}
\title{\vspace{-0.5em} \LARGE Two-Timescale Design for Movable Antenna Array-Enabled \\ Multiuser Uplink Communications}
\author{Guojie Hu, Qingqing Wu,~\textit{Senior Member}, \textit{IEEE}, Donghui Xu, Kui Xu,~\textit{Member}, \textit{IEEE}, Jiangbo Si,~\textit{Senior Member}, \textit{IEEE}, Yunlong Cai,~\textit{Senior Member}, \textit{IEEE}, and Naofal Al-Dhahir,~\textit{Fellow}, \textit{IEEE}\vspace{-2.8em}
 %and Yunlong Cai,~\textit{Senior Member}, \textit{IEEE}
\thanks{
%Copyright (c) 2015 IEEE. Personal use of this material is permitted. However, permission to use this material for any other purposes must be obtained from the IEEE by sending a request to pubs-permissions@ieee.org.
%
%This work was supported in part by the Natural Science Foundations
%of China under Grants 62201606, 62071485, 62271503, 62071480 and 61971337, and in part by Natural Science Foundation of Jiangsu Province under Grant BK 20201334 and Shaanxi Province Natural Science Foundation
%for Distinguished Young Scholar (2022JC-50). (Corresponding author: Jiangbo Si)
%This work was supported in part by the Natural Science Foundations of China under Grants 62201606.
%for Distinguished Young Scholar under Grant 61825104. (Corresponding author: Jiangbo Si)
Guojie Hu and Donghui Xu are with the College of Communication Engineering, Rocket Force University of Engineering, Xi'an 710025, China (lgdxhgj@sina.com). Qingqing Wu is with the Department of Electronic Engineering, Shanghai Jiao Tong University, Shanghai 200240, China. Kui Xu is with the College of Communications Engineering, Army Engineering University of PLA, Nanjing 210007, China. Jiangbo Si is with the Integrated Service Networks Lab of Xidian University, Xi'an 710100, China. Yunlong Cai is with the College of Information Science and Electronic Engineering, Zhejiang University, Hangzhou 310027, China. Naofal Al-Dhahir is with the Department of Electrical and Computer Engineering, The University of Texas at Dallas, Richardson, TX 75080 USA.
%Guojie Hu and Donghui Xu are with the College of Communication Engineering, Rocket Force University of Engineering, Xi'an 710025, China (email: lgdxhgj@sina.com). Jiangbo Si and Zan Li are with the Integrated Service Networks Lab of Xidian University, Xi'an 710100, China (e-mail: jbsi@xidian.edu.cn, zanli@xidian.edu.cn). Yunlong Cai is with the College of Information Science and Electronic Engineering, Zhejiang University, Hangzhou 310027, China (email: ylcai@zju.edu.cn). Naofal Al-Dhahir is with the Department of Electrical and Computer Engineering, The University of Texas at Dallas, Richardson, TX 75080 USA (e-mail: aldhahir@utdallas.edu).
}%\vspace{-1.4em}
}
%This work was supported by the Natural Science Foundations of China (No. 61671474).
%L. X. Yang, D. Wu, and Y. M. Cai are with the College of Communications Engineering, the Army of Engineering University, Nanjing 210007, China. (Email: yanglianxin1228@126.com; wujing1958725@126.com; caiym@vip.sina.com.
\IEEEpeerreviewmaketitle
%\vspace{-5pt}
\maketitle
%\vspace{-20pt}
\begin{abstract}
Movable antenna (MA) technology can flexibly reconfigure wireless channels by adjusting antenna positions in a local region, thus owing great potential for enhancing communication performance. This letter investigates MA technology-enabled multiuser uplink communications over general Rician fading channels, which consist of a base station (BS) equipped with the MA array and multiple single-antenna users. Since it is practically challenging to collect all instantaneous channel state information (CSI) by traversing all possible antenna positions at the BS, we instead propose a two-timescale scheme for maximizing the ergodic sum rate. Specifically, antenna positions at the BS are first optimized using only the statistical CSI. Subsequently, the receiving beamforming at the BS (for which we consider the three typical zero-forcing (ZF), minimum mean-square error (MMSE) and MMSE with successive interference cancellation (MMSE-SIC) receivers) is designed based on the instantaneous CSI with optimized antenna positions, thus significantly reducing practical implementation complexities. The formulated problems are highly non-convex and we develop projected gradient ascent (PGA) algorithms to effectively handle them. Simulation results illustrate that compared to conventional fixed-position antenna (FPA) array, the MA array can achieve significant performance gains by reaping an additional spatial degree of freedom.
\end{abstract}

\IEEEpeerreviewmaketitle
\vspace{-15pt}
\section{Introduction}
%\vspace{-5pt}
The multiuser uplink is an important communication scenario, which allows multiple users to send different signals in the same frequency to the base station (BS), which is usually equipped with multiple antennas for effectively decoding these signals with advanced receiving beamforming.

In conventional uplink communication, the BS adopts the fixed-position antenna (FPA) array \cite{Overview_MIMO}, where the distance between any two adjacent antennas is fixed. This setting may be undesirable, especially for dealing with some extreme scenarios. For instance, if there are some users located in close proximity with strong channel correlations, the BS with the FPA array may not be able to distinguish the corresponding signals well which may cause certain performance loss \cite{Guojie_TMC}.

Fortunately, the recent emerging movable antenna (MA) technology is expected to solve the above drawback. Specifically, unlike FPA, adopting MA at the BS, antenna positions can be flexibly adjusted in a specified region using stepper motors or servos \cite{Zhulipeng_CM,Flexible_MIMO_WCM}. Hence, the MA technology can effectively reconfigure wireless channels, provide an additional spatial degree of freedom (DoF) for varying channel correlations between different users, and thus create favorable conditions for implementing receiving beamforming \cite{Lipeng_Multiuser}.

Motivated by the significant potential of MA, some works have applied this technology to numerous scenarios, such as multiple-input multiple-output (MIMO) systems \cite{Mawenyan_MIMO}, physical-layer security networks \cite{Guojie_SPL}, spectrum sharing systems \cite{Weidong_SS}, interference networks \cite{Wang_HH_WCL}, multicast systems \cite{2024_Ying_MA}, multiuser downlink communications \cite{Qinhaoran_FAS_Downlink}, integrated sensing and communication (ISAC) \cite{Cunhua_MA_ISAC} and others \cite{Shuguang_MA_UAV}. In these works, one common characteristic is that they all considered multi-path sparse channels, where the number of paths is finite. Under this setting, the completed channel state information (CSI) can be easily obtained by estimating few parameters such as angles of departure (AoDs)/angles of arrival (AoAs) and complex coefficients of finite paths. Different from the above works, we in this paper consider a more challenging scenario where wireless channels suffer from the general Rician fading, for which the number of line-of-sight (LoS) paths between the BS and each user is only one, but the number of non-LoS (NLoS) paths approaches infinity. This feature of Rician fading imposes a fundamental difficulty, i.e., for each BS-user channel, its instantaneous and random NLoS component corresponding to different patterns of antenna positions at the BS may be different and then it is very costly for the BS to acquire the global instantaneous CSI by traversing all possible antenna positions. Facing this challenge, we propose a two-timescale scheme to maximize the ergodic sum rate exploiting the MA array. In detail, antenna positions at the BS are first optimized based on the statistical CSI, which is relatively static and thus can be easily estimated. Once antenna positions are optimized and fixed, the receiving beamforming at the BS is implemented with the instantaneous CSI, which can be obtained via classical channel estimation methods. Although the formulated problems are highly non-convex, we develop effective projected gradient ascent (PGA) methods to solve them. Our simulations show the considerable performance gains of the MA array.

%Current beamforming is implemented by leveraging fixed-position antennas (FPAs), where antennas cannot adjust their positions. This setting, undeniably, brings a fundamental limitation. Specifically, the design of beamforming is significantly related to wireless channels, which, however, cannot be reconfigured with FPAs. Then, if channel conditions are not desirable, beamforming with FPAs may not be able to attain its full potential \cite{SPL}. Motivated by this observation, one natural question arises, namely., whether wireless channels can be reconfigured to cater to effective beamforming designs? Recently, a promising technique called movable antennas (MAs) may provide the answer \cite{Zhulipeng_CM, FA1}. Using MAs, antenna positions at transmitters/receivers can be adjusted in a local region using the stepper motors or servos. This flexible behavior of MAs reshapes wireless channels adaptively, and thus brings an additional spatial degree of freedom (DoF) for further improving the system performance.

 \begin{figure}
 %\vspace{-10pt}
\centering
\includegraphics[width=5.5cm]{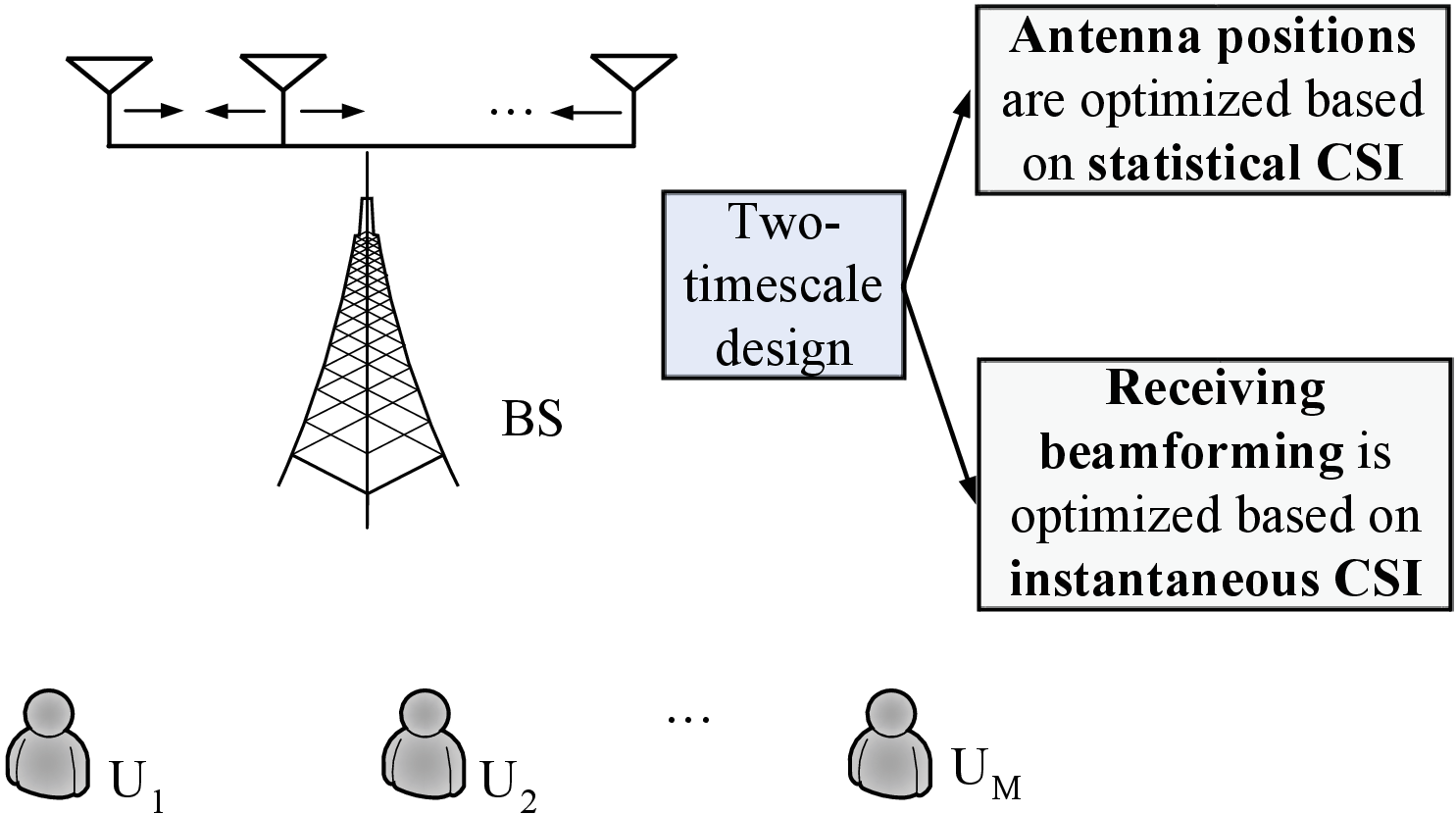}
\captionsetup{font=small}
\caption{Illustration of the system model.} \label{fig:Fig1}
\vspace{-15pt}
\end{figure}

 \newcounter{mytempeqncnt}
 %\vspace{-5pt}
\section{System Model and Problem Formulation}
%\vspace{-5pt}
\subsection{System Model}
As shown in Fig. 1, we consider the MA array-enabled multiuser uplink communications, where $M$ single-antenna users $\left\{ {{{\rm{U}}_m}} \right\}_{m = 1}^M$ transmit their messages to the MA array-empowered BS with $N$ antennas, and the one-dimensional positions of these $N$ MAs relative to the reference point zero are denoted as ${[{x_1},{x_2},...,{x_N}]^T} \buildrel \Delta \over = {\bf{x}}$. The equivalent baseband channel from ${\rm{U}}_m$ to the BS is denoted by ${{\bf{h}}_m}({\bf{x}}) \in {{\mathbb{C}}^{N \times 1}}$, $m = 1,2,...,M$, which is characterized with the general Rician fading model, i.e.,
\setlength\abovedisplayskip{1.4pt}
\setlength\belowdisplayskip{1.4pt}
\begin{equation}
{{\bf{h}}_m}({\bf{x}}) = \sqrt {\frac{{{K_m}{\beta _m}}}{{{K_m} + 1}}} {\overline {\bf{h}} _m}({\bf{x}}) + \sqrt {\frac{{{\beta _m}}}{{{K_m} + 1}}} {\widetilde {\bf{h}}_m}({\bf{x}}),
 \end{equation}
 where $K_m > 0$ is the Rician K-factor, ${{\beta _m}}$ is the large-scale fading coefficient, ${\overline {\bf{h}} _m}({\bf{x}}) \in {{\mathbb{C}}^{N \times 1}}$ is the LoS component, which is expressed as
 \begin{equation}
 {\overline {\bf{h}} _m}({\bf{x}}) = {\left[ {{e^{j\frac{{2\pi }}{\lambda }{x_1}\sin {\theta _m}}},...,{e^{j\frac{{2\pi }}{\lambda }{x_N}\sin {\theta _m}}}} \right]^T},
 \end{equation}
 where $\lambda $ is the carrier wavelength and ${{\theta _m}}$ is the arrival angle of ${\rm{U}}_m$. In addition, ${\widetilde {\bf{h}}_m}({\bf{x}}) \in {{\mathbb{C}}^{N \times 1}}$ in (1) is the NLoS component, where entries of ${\widetilde {\bf{h}}_m}({\bf{x}})$ are independent and identically distributed (i.i.d) circularly symmetric complex Gaussian random variables with zero mean and unit variance.

 Based on the above analysis, the received signal ${\bf{y}} \in {{\mathbb{C}}^{N \times 1}}$ at the BS can be expressed as
  \begin{equation}
 {\bf{y}} = {{\bf{h}}_m}({\bf{x}})\sqrt {{P_m}} {s_m} + \sum\nolimits_{i = 1,i \ne m}^M {{{\bf{h}}_i}({\bf{x}})\sqrt {{P_i}} {s_i}}  + {\bf{n}},
   \end{equation}
 where $P_m$ and $s_m$ are the transmit power and information-bearing signal of ${\rm{U}}_m$, and ${\bf{n}} \sim {\cal CN}\left( {{\bf{0}},{\sigma ^2}{{\bf{I}}_N}} \right)$ is the additive white Gaussian noise at the BS, with ${{\sigma ^2}}$ being the average noise power. Applying the receiving beamforming vector ${{\bf{w}}_m} \in {{\mathbb{C}}^{N \times 1}}$ for detecting ${\rm{U}}_m$'s signal, with $\left\| {{{\bf{w}}_m}} \right\| = 1$, the corresponding received signal-to-interference-plus-noise ratio (SINR) is
  \begin{equation}
 {\gamma _m} = \frac{{{{\overline P }_m}{{\left| {{\bf{w}}_m^H{{\bf{h}}_m}({\bf{x}})} \right|}^2}}}{{\sum\nolimits_{i = 1,i \ne m}^M {{{\overline P }_i}{{\left| {{\bf{w}}_m^H{{\bf{h}}_i}({\bf{x}})} \right|}^2}}  + 1}},
 \end{equation}
 where ${\overline P _m} = {P_m}/{\sigma ^2}$. Then, the achievable sum rate of all users in bits/s/Hz is expressed as
 \begin{equation}
 {R_{{\rm{sum}}}} = \sum\nolimits_{m = 1}^M {{{\log }_2}(1 + {\gamma _m})} .
 \end{equation}

 \subsection{Transmission Scheme}
 For the Rician fading channel ${{{\bf{h}}_m}({\bf{x}})}$, $m = 1,2,...,M$, note that its NLoS component ${\widetilde {\bf{h}}_m}({\bf{x}})$ corresponding to different pattern of antenna positions at the BS is totally random. This implies that to estimate all $\left\{ {{{\widetilde {\bf{h}}}_m}({\bf{x}})} \right\}_{m = 1}^M$ for arbitrary ${\bf{x}}$, it is necessary to traverse all possible antenna positions, which, however, is obviously infeasible due to exceedingly high complexity. Motivated by this observation, we propose a hierarchical transmission scheme. Specifically, in light of the fact that the LoS component of the Rician fading is relatively static, we consider that antenna positions at the BS are first optimized based on such statistical CSI \cite{yuanwei_timescale}. Once antenna positions are optimized and fixed, the BS can exploit classical channel estimation methods to acquire the instantaneous CSI of all uplink channels. Subsequently, the BS can rely on such instantaneous CSI to design its receiving beamforming.

 %\textbf{Remark:} Since the optimization of antenna positions is given one shot, and more importantly, channel estimations are only completed given one deterministic pattern of antenna positions, our proposed two-timescale scheme owns very low implementation complexity.

 \subsection{Problem Formulation}
Under the proposed transmission scheme, we aim to maximize the ergodic sum rate of all users by jointly optimizing long-term antenna positions ${\bf{x}}$ and short-term receiving beamforming vectors at the BS. Therefore, the problem can be formulated as
 \begin{align}
&({\rm{P1}}): \mathop {\max }\limits_{\bf{x}} {\mathbb{E}}\left[ {\mathop {\max }\limits_{\left\{ {{{\bf{w}}_m}} \right\}_{m = 1}^M} \sum\nolimits_{m = 1}^M {{{\log }_2}(1 + {\gamma _m})} } \right] \tag{${\rm{6a}}$}\\
{\rm{              }}&\ {\rm{s.t.}} \ \ \ \left\| {{{\bf{w}}_m}} \right\| = 1,\forall m = 1,...,M,\tag{${\rm{6b}}$}\\
 &\quad \quad \ \ \ {x_n} - {x_{n - 1}} \ge {d_{\min }},n = 2,3,...,N,\tag{${\rm{6c}}$}\\
&\quad \quad \ \ \left\{ {{x_n}} \right\}_{n = 1}^N \in [0,L],\tag{${\rm{6d}}$}
%& \ \ \ \quad {1_{\mathbb{C}}} = 1,\ {\rm{if}} \ \\
%&0 < {Q_E} \le \max \left( {{Q_{th}}/\left( {\frac{1}{{{\lambda _{EE}}}} + \frac{{{\beta %^r}N}}{{{\lambda _{RE}}{\lambda _{ER}}}}} \right),{Q_{E,\max }}} \right)\tag{${\rm{11d}}$}.
 \end{align}
where ${\mathbb{E}}\left[  \cdot  \right]$ in (6a) is the expectation of the sum rate by taking all channels' random realizations, $d_{\rm{min}}$ in (6c) is the minimum distance between any two adjacent MAs for avoiding the coupling effect, and $L$ in (6d) is the total antenna span.

Generally, (P1) is difficult to solve since i) the variables ${\left\{ {{{\bf{w}}_m}} \right\}_{m = 1}^M}$ and ${\bf{x}}$ are coupled with each other in the objective; ii) there is no closed-form expression for the ergodic sum rate even given ${\bf{x}}$. In the next section, we respectively investigate three classical receiving beamforming schemes, i.e., zero-forcing (ZF), minimum mean-square error (MMSE) and MMSE with successive interference cancellation (MMSE-SIC), for (P1). For each receiving scheme, we further design the effective PGA algorithm to optimize antenna positions.

 \begin{figure*}[b!]
% \vspace{-8pt}
  \hrulefill
\setcounter{mytempeqncnt}{\value{equation}}
\begin{equation}
\setcounter{equation}{12}
\begin{split}{}
\frac{{\partial {f_{{\rm{ZF}}}}({\bf{x}})}}{{\partial {x_n}}} = \frac{1}{{\ln 2}}\sum\nolimits_{m = 1}^M {\frac{{ - {{\overline P }_m}{\beta _m}(N - M)}}{{{{\left( {{{\left[ {{{\bf{\Sigma }}^{ - 1}}({\bf{x}})} \right]}_{mm}}} \right)}^2} + {{\overline P }_m}{\beta _m}(N - M){{\left[ {{{\bf{\Sigma }}^{ - 1}}({\bf{x}})} \right]}_{mm}}}}\frac{{\partial {{\left[ {{{\bf{\Sigma }}^{ - 1}}({\bf{x}})} \right]}_{mm}}}}{{\partial {x_n}}}}.
\end{split}
\end{equation}
\setcounter{equation}{\value{mytempeqncnt}}
%\vspace{-12pt}
\end{figure*}

\section{Solutions to (P1)}
\subsection{ZF Receiver}
Applying the ZF receiver, the inter-user interference (IUI) can be completely removed. Based on \cite{Zengyong_ZF_MMSE}, the ZF receiving beamforming for detecting ${\rm{U}}_m$'s signal is expressed as
\begin{equation} \small
\setcounter{equation}{7}
{{\bf{w}}_{{\rm{ZF}},m}} = \frac{{\left( {{{\bf{I}}_N} - {{\bf{A}}_m}({\bf{x}}){{({\bf{A}}_m^H({\bf{x}}){{\bf{A}}_m}({\bf{x}}))}^{ - 1}}{\bf{A}}_m^H({\bf{x}})} \right){{\bf{h}}_m}({\bf{x}})}}{{\left\| {\left( {{{\bf{I}}_N} - {{\bf{A}}_m}({\bf{x}}){{({\bf{A}}_m^H({\bf{x}}){{\bf{A}}_m}({\bf{x}}))}^{ - 1}}{\bf{A}}_m^H({\bf{x}})} \right){{\bf{h}}_m}({\bf{x}})} \right\|}},
\end{equation}
where ${{\bf{A}}_m}({\bf{x}}) = [{{\bf{h}}_1}({\bf{x}}),...,{{\bf{h}}_{m - 1}}({\bf{x}}),{{\bf{h}}_{m + 1}}({\bf{x}}),...,{{\bf{h}}_M}({\bf{x}})] \in {{\mathbb{C}}^{N \times (M - 1)}}$. Substituting (7) into (4), the received signal-to-noise ratio (SNR) for ${\rm{U}}_m$'s signal is \cite{Jinshi_JSTSP}
\begin{equation}
{\gamma _{{\rm{ZF}},m}} = {{\bar P}_m} \big /{\left[ {{{\left( {{{\bf{H}}^H}({\bf{x}}){\bf{H}}({\bf{x}})} \right)}^{ - 1}}} \right]_{mm}},
\end{equation}
where ${\bf{H}}({\bf{x}}) = [{{\bf{h}}_1}({\bf{x}}),{{\bf{h}}_2}({\bf{x}}),...,{{\bf{h}}_M}({\bf{x}})] \in {{\mathbb{C}}^{N \times M}}$. Based on (8) and given ${\bf{x}}$, the ergodic rate of ${\rm{U}}_m$ is $R_{{\rm{ZF}},m}^{{\rm{erg}}} = {\mathbb{E}}\left[ {{{\log }_2}\left( {1 + \frac{{{{\overline P }_m}}}{{{{\left[ {{{\left( {{{\bf{H}}^H}({\bf{x}}){\bf{H}}({\bf{x}})} \right)}^{ - 1}}} \right]}_{mm}}}}} \right)} \right]$. Note that the exact expression of $R_{{\rm{ZF}},m}^{{\rm{erg}}}$ is very hard to derive. To tackle this challenge, based on the conclusion in \cite{Jinshi_JSTSP}, we can obtain a tight approximation of $R_{{\rm{ZF}},m}^{{\rm{erg}}}$ as
\begin{equation}
\begin{split}{}
R_{{\rm{ZF}},m}^{{\rm{erg}}} \approx {\log _2}\left( {1 + \frac{{{{\overline P }_m}{\beta _m}(N - M)}}{{{{\left[ {{{\bf{\Sigma }}^{ - 1}}({\bf{x}})} \right]}_{mm}}}}} \right),
\end{split}
\end{equation}
where ${\bf{\Sigma }}({\bf{x}}) = {{\bf{\Theta }}_1} + \frac{1}{N}{{\bf{\Theta }}_2}{\overline {\bf{H}} ^H}({\bf{x}})\overline {\bf{H}} ({\bf{x}}){{\bf{\Theta }}_2}$, ${{\bf{\Theta }}_1} = {\left( {{\bf{\Omega }} + {{\bf{I}}_M}} \right)^{ - 1}} \in {{\mathbb{R}}^{M \times M}}$ with ${\bf{\Omega }} = {\rm{diag}}\left\{ {{K_1},{K_2},...,{K_M}} \right\} \in {{\mathbb{R}}^{M \times M}}$, ${{\bf{\Theta }}_2} = {\left[ {{\bf{\Omega }}{{\bf{\Theta }}_1}} \right]^{1/2}} \in {{\mathbb{R}}^{M \times M}}$, and $\overline {\bf{H}} ({\bf{x}}) = \left[ {{{\overline {\bf{h}} }_1}({\bf{x}}),{{\overline {\bf{h}} }_2}({\bf{x}}),...,{{\overline {\bf{h}} }_M}({\bf{x}})} \right] \in {{\mathbb{C}}^{N \times M}}$.

Based on (9), problem (P1) can be simplified as
 \begin{align}
&({\rm{P2}}): \mathop {\max }\limits_{\bf{x}} \sum\nolimits_{m = 1}^M {{{\log }_2}\left( {1 + \frac{{{{\overline P }_m}{\beta _m}(N - M)}}{{{{\left[ {{{\bf{\Sigma }}^{ - 1}}({\bf{x}})} \right]}_{mm}}}}} \right)}   \tag{${\rm{10a}}$}\\
{\rm{              }}&\ {\rm{s.t.}} \quad (6{\rm{c}}),(6{\rm{d}}). \tag{${\rm{10b}}$}
%& \ \ \ \quad {1_{\mathbb{C}}} = 1,\ {\rm{if}} \ \\
%&0 < {Q_E} \le \max \left( {{Q_{th}}/\left( {\frac{1}{{{\lambda _{EE}}}} + \frac{{{\beta %^r}N}}{{{\lambda _{RE}}{\lambda _{ER}}}}} \right),{Q_{E,\max }}} \right)\tag{${\rm{11d}}$}.
 \end{align}

Observing (10a), the objective of (P2) is still highly non-convex. To address this challenge, we next exploit the PGA algorithm to find a locally optimal solution to (P2). In detail, based on PGA, the update procedure for ${\bf{x}}$ in the $t + 1$-th iteration is given by
\begin{equation}
\setcounter{equation}{11}
\begin{split}{}
{{\bf{x}}^{t + 1}} =& {{\bf{x}}^t} + \delta {\nabla _{{{\bf{x}}^t}}}{f_{{\rm{ZF}}}}({\bf{x}}),\\
{{\bf{x}}^{t + 1}} =& {\cal B}\left\{ {{{\bf{x}}^{t + 1}},{\cal C}} \right\},
\end{split}
\end{equation}
where ${f_{{\rm{ZF}}}}({\bf{x}}) \buildrel \Delta \over = \sum\nolimits_{m = 1}^M {{{\log }_2}\left( {1 + \frac{{{{\overline P }_m}{\beta _m}(N - M)}}{{{{\left[ {{{\bf{\Sigma }}^{ - 1}}({\bf{x}})} \right]}_{mm}}}}} \right)} $, $\delta $ is the step size for the gradient ascent which can be determined via the backtracking line
search \cite{Lipeng_Multiuser}, ${\nabla _{{{\bf{x}}^t}}}{f_{{\rm{ZF}}}}({\bf{x}})$ is the gradient of ${f_{{\rm{ZF}}}}({\bf{x}})$ at ${{{\bf{x}}^t}}$, and ${\cal B}\left\{  \cdot  \right\}$ is the projection function (as shown later) which ensures that the updates for antenna positions in each iteration always lie in the feasible region specified by (6c).

\textit{Determining The Gradient:} The key of the proposed PGA algorithm is to derive the expression of ${\nabla _{{{\bf{x}}^t}}}{f_{{\rm{ZF}}}}({\bf{x}})$. Note that ${\nabla _{{{\bf{x}}^t}}}{f_{{\rm{ZF}}}}({\bf{x}}) = \left[ {\frac{{\partial {f_{{\rm{ZF}}}}({\bf{x}})}}{{\partial {x_1}}},...,\frac{{\partial {f_{{\rm{ZF}}}}({\bf{x}})}}{{\partial {x_N}}}} \right]_{{\bf{x}} = {{\bf{x}}^t}}^T$, where ${\frac{{\partial {f_{{\rm{ZF}}}}({\bf{x}})}}{{\partial {x_n}}}}$, $\forall n = 1,...,N$, can be derived as in (12), with
\begin{equation}
\setcounter{equation}{13}
\frac{{\partial {{\left[ {{{\bf{\Sigma }}^{ - 1}}({\bf{x}})} \right]}_{mm}}}}{{\partial {x_n}}} = {\left[ {\frac{{\partial {{\bf{\Sigma }}^{ - 1}}({\bf{x}})}}{{\partial {x_n}}}} \right]_{mm}},
\end{equation}
 and based on the inverse matrix differentiation law, we have
 \begin{equation}
\frac{{\partial {{\bf{\Sigma }}^{ - 1}}({\bf{x}})}}{{\partial {x_n}}} =  - {{\bf{\Sigma }}^{ - 1}}({\bf{x}})\frac{{\partial {\bf{\Sigma }}({\bf{x}})}}{{\partial {x_n}}}{{\bf{\Sigma }}^{ - 1}}({\bf{x}}),
\end{equation}
where
 \begin{equation}
\frac{{\partial {\bf{\Sigma }}({\bf{x}})}}{{\partial {x_n}}} = \frac{1}{N}{{\bf{\Theta }}_2}\left( {\frac{{\partial {{\overline {\bf{H}} }^H}({\bf{x}})}}{{\partial {x_n}}}\overline {\bf{H}} ({\bf{x}}) + {{\overline {\bf{H}} }^H}({\bf{x}})\frac{{\partial \overline {\bf{H}} ({\bf{x}})}}{{\partial {x_n}}}} \right){{\bf{\Theta }}_2},
\end{equation}
with $\frac{{\partial {{\overline {\bf{H}} }^H}({\bf{x}})}}{{\partial {x_n}}} = {\left[ {\frac{{\partial \overline {\bf{H}} ({\bf{x}})}}{{\partial {x_n}}}} \right]^H} \in {{\mathbb{C}}^{M \times N}}$ and
\begin{equation}
\frac{{\partial \overline {\bf{H}} ({\bf{x}})}}{{\partial {x_n}}} = \frac{{2\pi }}{\lambda }{\left[ {\begin{array}{*{20}{c}}
{{{\bf{0}}^{M \times 1}}}& \cdots &{{\bf{b}}({x_n})}& \cdots &{{{\bf{0}}^{M \times 1}}}
\end{array}} \right]^T},
\end{equation}
with ${\bf{b}}({x_n}) = {\left[ {\begin{array}{*{20}{c}}
{\sin {\theta _1}{e^{j\left( {\frac{{2\pi }}{\lambda }{x_n}\sin {\theta _1} + \frac{\pi }{2}} \right)}}}\\
 \vdots \\
{\sin {\theta _M}{e^{j\left( {\frac{{2\pi }}{\lambda }{x_n}\sin {\theta _M} + \frac{\pi }{2}} \right)}}}
\end{array}} \right]}$.

Via the above analysis, substituting (14) into (13) and then substituting (13) into (12), ${\nabla _{{{\bf{x}}^t}}}{f_{{\rm{ZF}}}}({\bf{x}})$ can be finally obtained.

\textit{Determining The Projected Function:} Since the updates for antenna positions in each iteration should strictly satisfy the constraint in (6c), according to the nearest distance rule explained in our previous work \cite{Guojie_SPL}, the projection function in (11) can be obtained as
\begin{equation} \nonumber
\begin{split}{}
&{\cal B}\left\{ {{{\bf{x}}^{t + 1}},{\cal C}} \right\}:\\
&\left\{ {\begin{array}{*{20}{c}}
{x_1^{t + 1} = \max \left( {0,\min \left( {L - (N - 1){d_{\min }},x_1^{t + 1}} \right)} \right),}\\
{x_2^{t + 1} = \max \left( {x_1^{t + 1} + {d_{\min }},\min \left( {L - (N - 2){d_{\min }},x_2^{t + 1}} \right)} \right),}\\
{...}\\
{x_N^{t + 1} = \max \left( {x_{N - 1}^{t + 1} + {d_{\min }},\min \left( {L,x_N^{t + 1}} \right)} \right).}
\end{array}} \right.
\end{split}
\end{equation}

 \begin{figure*}[b!]
% \vspace{-2pt}
  \hrulefill
\setcounter{mytempeqncnt}{\value{equation}}
\begin{equation}
\setcounter{equation}{21}
\begin{split}{}
\frac{{\partial {f_{{\rm{MMSE}}}}({\bf{x}})}}{{\partial {x_n}}} = \frac{1}{{S\ln 2}}\sum\nolimits_{m = 1}^M {\sum\nolimits_{s = 1}^S {\frac{{{{\bar P}_m}}}{{1 + {{\bar P}_m}{\bf{h}}_{m,s}^H({\bf{x}}){\bf{B}}_{m,s}^{ - 1}({\bf{x}}){{\bf{h}}_{m,s}}({\bf{x}})}}\frac{{\partial \left[ {{\bf{h}}_{m,s}^H({\bf{x}}){\bf{B}}_{m,s}^{ - 1}({\bf{x}}){{\bf{h}}_{m,s}}({\bf{x}})} \right]}}{{\partial {x_n}}}} }  .
\end{split}
\end{equation}
\setcounter{equation}{\value{mytempeqncnt}}
%\vspace{-2pt}
\end{figure*}

%  \begin{figure*}[b!]
% %\vspace{-2pt}
%  \hrulefill
%\setcounter{mytempeqncnt}{\value{equation}}
%\begin{equation}
%\setcounter{equation}{30}
%\begin{split}{}
%&{\mathbb{E}}\left[ {{{\bf{h}}_m}({\bf{x}}){\bf{h}}_m^H({\bf{x}})} \right] = {\mathbb{E}}\left\{ {\left[ {\sqrt {\frac{{{K_m}{\beta _m}}}{{{K_m} + 1}}} {{\overline {\bf{h}} }_m}({\bf{x}}) + \sqrt {\frac{{{\beta _m}}}{{{K_m} + 1}}} {{\widetilde {\bf{h}}}_m}({\bf{x}})} \right]{{\left[ {\sqrt {\frac{{{K_m}{\beta _m}}}{{{K_m} + 1}}} {{\overline {\bf{h}} }_m}({\bf{x}}) + \sqrt {\frac{{{\beta _m}}}{{{K_m} + 1}}} {{\widetilde {\bf{h}}}_m}({\bf{x}})} \right]}^H}} \right\}\\
% =& {\mathbb{E}}\left\{ {\left[ {\frac{{{K_m}{\beta _m}}}{{{K_m} + 1}}{{\overline {\bf{h}} }_m}({\bf{x}})\overline {\bf{h}} _m^H({\bf{x}}) + \frac{{{\beta _m}\sqrt {{K_m}} }}{{{K_m} + 1}}{{\overline {\bf{h}} }_m}({\bf{x}})\widetilde {\bf{h}}_m^H({\bf{x}}) + \frac{{{\beta _m}\sqrt {{K_m}} }}{{{K_m} + 1}}{{\widetilde {\bf{h}}}_m}({\bf{x}})\overline {\bf{h}} _m^H({\bf{x}}) + \frac{{{\beta _m}}}{{{K_m} + 1}}{{\widetilde {\bf{h}}}_m}({\bf{x}})\widetilde {\bf{h}}_m^H({\bf{x}})} \right]} \right\}\\
% =& \frac{{{K_m}{\beta _m}}}{{{K_m} + 1}}{\overline {\bf{h}} _m}({\bf{x}})\overline {\bf{h}} _m^H({\bf{x}}) + \frac{{{\beta _m}}}{{{K_m} + 1}}{{\bf{I}}_N}.
%\end{split}
%\end{equation}
%\setcounter{equation}{\value{mytempeqncnt}}
%%\vspace{-2pt}
%\end{figure*}

  \begin{figure*}[b!]
% \vspace{-2pt}
  \hrulefill
\setcounter{mytempeqncnt}{\value{equation}}
\begin{equation}
\setcounter{equation}{31}
\begin{split}{}
\frac{{\partial {f_{{\rm{MMSE - SIC}}}}({\bf{x}})}}{{\partial {x_n}}} = \frac{1}{{\ln 2}}{\rm{Tr}}\left\{ {{\bf{D}}({\bf{x}})\left( {\sum\nolimits_{m = 1}^M {\frac{{{{\overline P }_m}{K_m}{\beta _m}}}{{{K_m} + 1}}\left( {\frac{{\partial {{\overline {\bf{h}} }_m}({\bf{x}})}}{{\partial {x_n}}}\overline {\bf{h}} _m^H({\bf{x}}) + {{\overline {\bf{h}} }_m}({\bf{x}})\frac{{\partial \overline {\bf{h}} _m^H({\bf{x}})}}{{\partial {x_n}}}} \right)} } \right)} \right\}.
\end{split}
\end{equation}
\setcounter{equation}{\value{mytempeqncnt}}
%\vspace{-2pt}
\end{figure*}

\textit{Complexity Analysis:} The PGA algorithm for solving (P2) starts by generalizing an initial ${{\bf{x}}^0}$ and then recursively updating antenna positions based on (11) until the objective of (P2) converges to a constant. To simplify the analysis while still keeping a good approximation, we now focus on the number of complex multiplications required per iteration. In detail, from (12)$-$(15), we can conclude that: the complexity of computing ${\frac{{\partial {{\overline {\bf{H}} }^H}({\bf{x}})}}{{\partial {x_n}}}\overline {\bf{H}} ({\bf{x}})}$ or ${{{\overline {\bf{H}} }^H}({\bf{x}})\frac{{\partial \overline {\bf{H}} ({\bf{x}})}}{{\partial {x_n}}}}$ is ${\cal O}(N{M^2})$, based on which the complexities of computing $\frac{{\partial {\bf{\Sigma }}({\bf{x}})}}{{\partial {x_n}}}$ and then $\frac{{\partial {{\bf{\Sigma }}^{ - 1}}({\bf{x}})}}{{\partial {x_n}}}$ can be, respectively, derived as ${\cal O}(2N{M^2} + 2{M^3})$ and ${\cal O}(2N{M^2} + 5{M^3})$, where we have exploited the fact that for a matrix ${\bf{\Sigma }}({\bf{x}}) \in {{\mathbb{C}}^{M \times M}}$, computing ${{{\bf{\Sigma }}^{ - 1}}({\bf{x}})}$ requires the complexity of ${\cal O}({M^3})$. Based on the above analysis, computing ${\nabla _{{{\bf{x}}^t}}}{f_{{\rm{ZF}}}}({\bf{x}})$ in each iteration requires the complexity of ${\cal O}(2{N^2}{M^2} + 5N{M^3})$, leading to the total complexity of obtaining a stationary point of (P2) as ${\cal O}\left( {{I_{{\rm{ite}}}}(2{N^2}{M^2} + 5N{M^3})} \right)$, where ${{I_{{\rm{ite}}}}}$ is the number of iterations.

\subsection{MMSE Receiver}
Compared to the ZF receiver, the MMSE receiver achieves a best tradeoff between maximizing the useful channel power gain and minimizing the IUI, thus maximizing the SINR in (4). Specifically, the MMSE receiving beamforming for detecting ${\rm{U}}_m$'s signal is expressed as \cite{Zengyong_ZF_MMSE}
\begin{equation}
{{\bf{w}}_{{\rm{MMSE}},m}} = {\bf{B}}_m^{ - 1}({\bf{x}}){{\bf{h}}_m}({\bf{x}})\big /\left\| {{\bf{B}}_m^{ - 1}({\bf{x}}){{\bf{h}}_m}({\bf{x}})} \right\|,
%\end{split}
\end{equation}
where ${{\bf{B}}_m}({\bf{x}}) = \sum\nolimits_{i = 1,i \ne m}^M {{{\overline P }_i}{{\bf{h}}_i}({\bf{x}}){\bf{h}}_i^H({\bf{x}}) + {{\bf{I}}_N}}  \in {{\mathbb{C}}^{N \times N}}$. Substituting (17) into (4), the received SINR for ${\rm{U}}_m$'s signal is \cite{Zengyong_ZF_MMSE}
\begin{equation}
\begin{split}
{\gamma _{{\rm{MMSE}},m}} = {\overline P _m}{\bf{h}}_m^H({\bf{x}}){\bf{B}}_m^{ - 1}({\bf{x}}){{\bf{h}}_m}({\bf{x}}).
\end{split}
\end{equation}
Therefore, based on (18) and given ${\bf{x}}$, the ergodic rate of ${\rm{U}}_m$ is $R_{{\rm{MMSE}},m}^{{\rm{erg}}} = {\mathbb{E}}\left[ {{{\log }_2}\left( {1 + {{\overline P }_m}{\bf{h}}_m^H({\bf{x}}){\bf{B}}_m^{ - 1}({\bf{x}}){{\bf{h}}_m}({\bf{x}})} \right)} \right]$. Unfortunately, it is difficult to derive the exact expression of $R_{{\rm{MMSE}},m}^{{\rm{erg}}}$ and even its tight approximation. This motivates us to employ the standard Monte Carlo method for obtaining an approximation of $R_{{\rm{MMSE}},m}^{{\rm{erg}}}$. Specifically, we first generate $S$ independent and random realizations for $\left\{ {{{\widetilde {\bf{h}}}_m}({\bf{x}})} \right\}_{m = 1}^M$ according to their distribution information, which are denoted as $\left\{ {\left\{ {{{\widetilde {\bf{h}}}_{m,s}}({\bf{x}})} \right\}_{m = 1}^M} \right\}_{s = 1}^S$. Then, $R_{{\rm{MMSE}},m}^{{\rm{erg}}}$ can be approximated as
\begin{equation}
\begin{split}{}
&R_{{\rm{MMSE}},m}^{{\rm{erg}},{\rm{appr}}}\\
 =& \frac{1}{S}\sum\nolimits_{s = 1}^S {{{\log }_2}\left( {1 + {{\overline P }_m}{\bf{h}}_{m,s}^H({\bf{x}}){\bf{B}}_{m,s}^{ - 1}({\bf{x}}){{\bf{h}}_{m,s}}({\bf{x}})} \right)},
\end{split}
\end{equation}
where ${{\bf{h}}_{m,s}}({\bf{x}}) = {{\bf{h}}_m}({\bf{x}}){|_{{{\widetilde {\bf{h}}}_m}({\bf{x}}) = {{\widetilde {\bf{h}}}_{m,s}}({\bf{x}})}}$ and ${\bf{B}}_{m,s}^{ - 1}({\bf{x}}) = {\bf{B}}_m^{ - 1}({\bf{x}}){|_{\left\{ {{{\widetilde {\bf{h}}}_i}({\bf{x}}) = {{\widetilde {\bf{h}}}_{i,s}}({\bf{x}})} \right\}_{i = 1,i \ne m}^M}}$.

Based on (19), problem (P1) can be simplified as
 \begin{align}
&({\rm{P3}}): \mathop {\max }\limits_{\bf{x}} \sum\nolimits_{m = 1}^M {R_{{\rm{MMSE}},m}^{{\rm{erg}},{\rm{appr}}}}    \tag{${\rm{20a}}$}\\
{\rm{              }}&\ {\rm{s.t.}} \quad (6{\rm{c}}),(6{\rm{d}}). \tag{${\rm{20b}}$}
%& \ \ \ \quad {1_{\mathbb{C}}} = 1,\ {\rm{if}} \ \\
%&0 < {Q_E} \le \max \left( {{Q_{th}}/\left( {\frac{1}{{{\lambda _{EE}}}} + \frac{{{\beta %^r}N}}{{{\lambda _{RE}}{\lambda _{ER}}}}} \right),{Q_{E,\max }}} \right)\tag{${\rm{11d}}$}.
 \end{align}

 Since (P3) is highly non-convex, we still employ the PGA algorithm to tackle it. Note that in each iteration, the general update procedure for antenna positions has been provided in Section III-A and thus the details are not repeated for brevity. The only difference is that with the MMSE receiver, the gradient ${\nabla _{{{\bf{x}}^t}}}{f_{{\rm{ZF}}}}({\bf{x}})$ in (11) should be replaced with ${\nabla _{{{\bf{x}}^t}}}{f_{{\rm{MMSE}}}}({\bf{x}})$, with ${f_{{\rm{MMSE}}}}({\bf{x}})  \buildrel \Delta \over =  \sum\nolimits_{m = 1}^M {R_{{\rm{MMSE}},m}^{{\rm{erg}},{\rm{appr}}}} $. Then, we can derive that ${\nabla _{{{\bf{x}}^t}}}{f_{{\rm{MMSE}}}}({\bf{x}}) = \left[ {\frac{{\partial {f_{{\rm{MMSE}}}}({\bf{x}})}}{{\partial {x_1}}},...,\frac{{\partial {f_{{\rm{MMSE}}}}({\bf{x}})}}{{\partial {x_N}}}} \right]_{{\bf{x}} = {{\bf{x}}^t}}^T$, where ${\frac{{\partial {f_{{\rm{MMSE}}}}({\bf{x}})}}{{\partial {x_n}}}}$, $\forall n = 1,...,N$, can be derived as in (21), with
 \begin{equation} \small
 \setcounter{equation}{22}
\begin{split}{}
&\frac{{\partial \left[ {{\bf{h}}_{m,s}^H({\bf{x}}){\bf{B}}_{m,s}^{ - 1}({\bf{x}}){{\bf{h}}_{m,s}}({\bf{x}})} \right]}}{{\partial {x_n}}} = \frac{{\partial {\bf{h}}_{m,s}^H({\bf{x}})}}{{\partial {x_n}}}{\bf{B}}_{m,s}^{ - 1}({\bf{x}}){{\bf{h}}_{m,s}}({\bf{x}})\\
 +& {\bf{h}}_{m,s}^H({\bf{x}})\frac{{\partial {\bf{B}}_{m,s}^{ - 1}({\bf{x}})}}{{\partial {x_n}}}{{\bf{h}}_{m,s}}({\bf{x}}) + {\bf{h}}_{m,s}^H({\bf{x}}){\bf{B}}_{m,s}^{ - 1}({\bf{x}})\frac{{\partial {{\bf{h}}_{m,s}}({\bf{x}})}}{{\partial {x_n}}},
\end{split}
\end{equation}
where
 \begin{equation}
 \setcounter{equation}{23}
\begin{split}{}
\frac{{\partial {{\bf{h}}_{m,s}}({\bf{x}})}}{{\partial {x_n}}} = {\left[ {\begin{array}{*{20}{c}}
0& \cdots &{{c_m}({x_n})}& \cdots &0
\end{array}} \right]^T},
 \end{split}
\end{equation}
with ${c_m}({x_n}) = \sqrt {\frac{{{K_m}{\beta _m}}}{{{K_m} + 1}}} \frac{{2\pi }}{\lambda }\sin {\theta _m}{e^{j\left( {\frac{{2\pi }}{\lambda }{x_n}\sin {\theta _m} + \frac{\pi }{2}} \right)}}$. In addition, $\frac{{\partial {\bf{h}}_{m,s}^H({\bf{x}})}}{{\partial {x_n}}} = {\left[ {\frac{{\partial {{\bf{h}}_{m,s}}({\bf{x}})}}{{\partial {x_n}}}} \right]^H}$ and
 \begin{equation}
\begin{split}{}
\frac{{\partial {\bf{B}}_{m,s}^{ - 1}({\bf{x}})}}{{\partial {x_n}}} =  - {\bf{B}}_{m,s}^{ - 1}({\bf{x}})\frac{{\partial {{\bf{B}}_{m,s}}({\bf{x}})}}{{\partial {x_n}}}{\bf{B}}_{m,s}^{ - 1}({\bf{x}}),
 \end{split}
\end{equation}
 where
 \begin{equation}
\begin{split}{}
\frac{{\partial {{\bf{B}}_{m,s}}({\bf{x}})}}{{\partial {x_n}}} =& \sum\nolimits_{i = 1,i \ne m}^M {{{\overline P }_i}} \\
 \times & \left[ {\frac{{\partial {{\bf{h}}_{i,s}}({\bf{x}})}}{{\partial {x_n}}}{\bf{h}}_{i,s}^H({\bf{x}}) + {{\bf{h}}_{i,s}}({\bf{x}})\frac{{\partial {\bf{h}}_{i,s}^H({\bf{x}})}}{{\partial {x_n}}}} \right].
 \end{split}
\end{equation}

Via the above analysis, substituting (23)$-$(25) into (22) and then substituting (22) into (21), ${\nabla _{{{\bf{x}}^t}}}{f_{{\rm{MMSE}}}}({\bf{x}})$ can be finally obtained.

\textit{Complexity Analysis:} Similarly, computing ${\bf{B}}_{m,s}^{ - 1}({\bf{x}})$ and then $\frac{{\partial {\bf{B}}_{m,s}^{ - 1}({\bf{x}})}}{{\partial {x_n}}}$ in (24) requires the complexities of ${\cal O}({N^3})$ and ${\cal O}(3{N^3})$, respectively. Therefore, computing $\frac{{\partial \left[ {{\bf{h}}_{m,s}^H({\bf{x}}){\bf{B}}_{m,s}^{ - 1}({\bf{x}}){{\bf{h}}_{m,s}}({\bf{x}})} \right]}}{{\partial {x_n}}}$ in (22) requires the complexity of ${\cal O}\left( {3({N^3} + {N^2} + N)} \right)$, based on which it is easy to derive the complexities of computing $\frac{{\partial {f_{{\rm{MMSE}}}}({\bf{x}})}}{{\partial {x_n}}}$ and ${\nabla _{{{\bf{x}}^t}}}{f_{{\rm{MMSE}}}}({\bf{x}})$ as ${\cal O}\left( {3MS({N^3} + {N^2} + N)} \right)$ and ${\cal O}\left( {3MS({N^4} + {N^3} + {N^2})} \right)$, respectively. Hence, the total complexity of obtaining a stationary value of (P3) is about ${\cal O}\left( {3{I_{{\rm{ite}}}}MS({N^4} + {N^3} + {N^2})} \right)$.

\subsection{MMSE-SIC Receiver}
Compared to the MMSE receiver, the MMSE-SIC receiver can further alleviate the IUI by adding the SIC technique at the BS \cite{Zengming_SIC}. However, this enhancement is at the cost of increasing the hardware complexity.

Without loss of generality, we consider that users' signals are decoded successively based on their indices, i.e., ${\rm{U}}_1$'s signal is decoded first, while ${\rm{U}}_M$'s signal is decoded last. Applying the SIC technique, the BS can completely remove the IUI from $\left\{ {{{\rm{U}}_i}} \right\}_{i = 1}^{m - 1}$ before decoding ${\rm{U}}_m$'s signal. Therefore, the corresponding SINR for ${\rm{U}}_m$'s signal is
 \begin{equation}
\begin{split}{}
{\gamma _m} = \frac{{{{\overline P }_m}{{\left| {{\bf{w}}_m^H{{\bf{h}}_m}({\bf{x}})} \right|}^2}}}{{\sum\nolimits_{i = m + 1}^M {{{\overline P }_i}{{\left| {{\bf{w}}_m^H{{\bf{h}}_i}({\bf{x}})} \right|}^2}}  + 1}},
  \end{split}
\end{equation}
which can be maximized by adopting the MMSE-based receiving beamforming, i.e.,
 \begin{equation}
\begin{split}{}
{{\bf{w}}_{{\rm{MMSE - SIC}},m}} = {\bf{C}}_m^{ - 1}({\bf{x}}){{\bf{h}}_m}({\bf{x}})/\left\| {{\bf{C}}_m^{ - 1}({\bf{x}}){{\bf{h}}_m}({\bf{x}})} \right\|,
  \end{split}
\end{equation}
with ${{\bf{C}}_m}({\bf{x}}) = \sum\nolimits_{i = m + 1}^M {{{\bar P}_i}{{\bf{h}}_i}({\bf{x}}){\bf{h}}_i^H({\bf{x}}) + {{\bf{I}}_N}}  \in {{\mathbb{C}}^{N \times N}}$. Substituting (27) into (26) and based on the conclusion in \cite{Zengming_SIC}, we can directly derive the sum rate of all users as
\begin{equation}
\begin{split}{}
{R_{{\rm{sum}}}} = {\log _2}\det \left( {{{\bf{I}}_N} + \sum\nolimits_{m = 1}^M {{{\overline P }_m}{{\bf{h}}_m}({\bf{x}}){\bf{h}}_m^H({\bf{x}})} } \right).
\end{split}
\end{equation}

%\vspace{-5pt}
%\begin{figure*}
%%\vspace{-10pt}
%\centering
%
%\begin{minipage}{5.5cm}
%\includegraphics[width=5.5cm]{beam_gain.eps}
%\centering
%\vspace{-8pt}
%\subfigure{(a)}
%
%\end{minipage}
%\begin{minipage}{5.55cm}
%\includegraphics[width=5.6cm]{ver_N.eps}
%\centering
%\vspace{-8pt}
%\subfigure{(b)}
%
%\end{minipage}
%\begin{minipage}{5.55cm}
%\includegraphics[width=5.6cm]{ver_L.eps}
%\centering
%\vspace{-8pt}
%\subfigure{(c)}
%
%\end{minipage}
%%\vspace{-3pt}
%\captionsetup{font=small}
%\caption{(a) Beam gain of four schemes at different AoDs, $N = 5$ and $L = 4$; (b) Minimum rate versus the number of antennas at $\rm{S}$ ($N$), $L = 5$; (c) Minimum rate versus the total span of movable antennas at $\rm{S}$ ($L$), $N= 5$.}
%\vspace{-20pt}
%\end{figure*}

Although the exact expression of the ergodic ${R_{{\rm{sum}}}}$ is hard to obtain, we can exploit the Jensen's inequality to derive its upper bound, i.e.,
\begin{equation}
\begin{split}{}
&{\mathbb{E}}\left[ {{R_{{\rm{sum}}}}} \right] \le {\widehat R_{{\rm{sum}}}}\\
 =& {\log _2}\det \left( {{{\bf{I}}_N} + \sum\nolimits_{m = 1}^M {{{\overline P }_m}{\mathbb{E}}\left[ {{{\bf{h}}_m}({\bf{x}}){\bf{h}}_m^H({\bf{x}})} \right]} } \right),
\end{split}
\end{equation}
where ${{\mathbb{E}}\left[ {{{\bf{h}}_m}({\bf{x}}){\bf{h}}_m^H({\bf{x}})} \right]}$ after some simple manipulations can be derived as ${\mathbb{E}}\left[ {{{\bf{h}}_m}({\bf{x}}){\bf{h}}_m^H({\bf{x}})} \right] = \frac{{{K_m}{\beta _m}}}{{{K_m} + 1}}{\overline {\bf{h}} _m}({\bf{x}})\overline {\bf{h}} _m^H({\bf{x}}) + \frac{{{\beta _m}}}{{{K_m} + 1}}{{\bf{I}}_N}$. Based on (29), we can formulate the problem as
 \begin{align}
&({\rm{P4}}): \mathop {\max }\limits_{\bf{x}} {f_{{\rm{MMSE}} - {\rm{SIC}}}}({\bf{x}}) \buildrel \Delta \over = {\widehat R_{{\rm{sum}}}}   \tag{${\rm{30a}}$}\\
{\rm{              }}&\ {\rm{s.t.}} \quad (6{\rm{c}}),(6{\rm{d}}). \tag{${\rm{30b}}$}
%& \ \ \ \quad {1_{\mathbb{C}}} = 1,\ {\rm{if}} \ \\
%&0 < {Q_E} \le \max \left( {{Q_{th}}/\left( {\frac{1}{{{\lambda _{EE}}}} + \frac{{{\beta %^r}N}}{{{\lambda _{RE}}{\lambda _{ER}}}}} \right),{Q_{E,\max }}} \right)\tag{${\rm{11d}}$}.
 \end{align}

Next, the PGA algorithm is applied to solve (P4). Similarly, we need to derive the gradient ${\nabla _{{{\bf{x}}^t}}}{f_{\rm{MMSE} - \rm{SIC}}}({\bf{x}})$ in each iteration. In detail, note that ${\nabla _{{{\bf{x}}^t}}}{f_{{\rm{MMSE - SIC}}}}({\bf{x}}) = \left[ {\frac{{\partial {f_{{\rm{MMSE - SIC}}}}({\bf{x}})}}{{\partial {x_1}}},...,\frac{{\partial {f_{{\rm{MMSE - SIC}}}}({\bf{x}})}}{{\partial {x_N}}}} \right]_{{\bf{x}} = {{\bf{x}}^t}}^T$, where ${\frac{{\partial {f_{{\rm{MMSE}}}}({\bf{x}})}}{{\partial {x_n}}}}$, $\forall n = 1,...,N$, can be derived as in (31), with ${\bf{D}}({\bf{x}}) = {\left( {{{\bf{I}}_N} + \sum\nolimits_{m = 1}^M {{{\overline P }_m}{\mathbb{E}}\left[ {{{\bf{h}}_m}({\bf{x}}){\bf{h}}_m^H({\bf{x}})} \right]} } \right)^{ - 1}}$, $\frac{{\partial {{\overline {\bf{h}} }_m}({\bf{x}})}}{{\partial {x_n}}} = {\left[ {\begin{array}{*{20}{c}}
0& \cdots &{d_m}({x_n})& \cdots &0
\end{array}} \right]^T}$, ${d_m}({x_n}) = \frac{{2\pi }}{\lambda }\sin {\theta _m}{e^{j\left( {\frac{{2\pi }}{\lambda }{x_n}\sin {\theta _m} + \frac{\pi }{2}} \right)}}$ and $\frac{{\partial \overline {\bf{h}} _m^H({\bf{x}})}}{{\partial {x_n}}} = {\left[ {\frac{{\partial {{\overline {\bf{h}} }_m}({\bf{x}})}}{{\partial {x_n}}}} \right]^H}$.

\textit{Complexity Analysis:} The complexity of computing ${\bf{D}}({\bf{x}})$ is about ${\cal O}({N^3} + M{N^2})$, based on which it is easy to derive the complexity of computing $\frac{{\partial {f_{{\rm{MMSE - SIC}}}}({\bf{x}})}}{{\partial {x_n}}}$ as ${\cal O}(2{N^3} + 3M{N^2})$. Then, the complexity of computing ${\nabla _{{{\bf{x}}^t}}}{f_{{\rm{MMSE - SIC}}}}({\bf{x}})$ is ${\cal O}(2{N^4} + 3M{N^3})$, leading to the total complexity of obtaining a stationary value of (P4) as ${\cal O}\left( {{I_{{\rm{ite}}}}(2{N^4} + 3M{N^3})} \right)$.

 \begin{figure}
 %\vspace{-10pt}
\centering
\includegraphics[width=6cm]{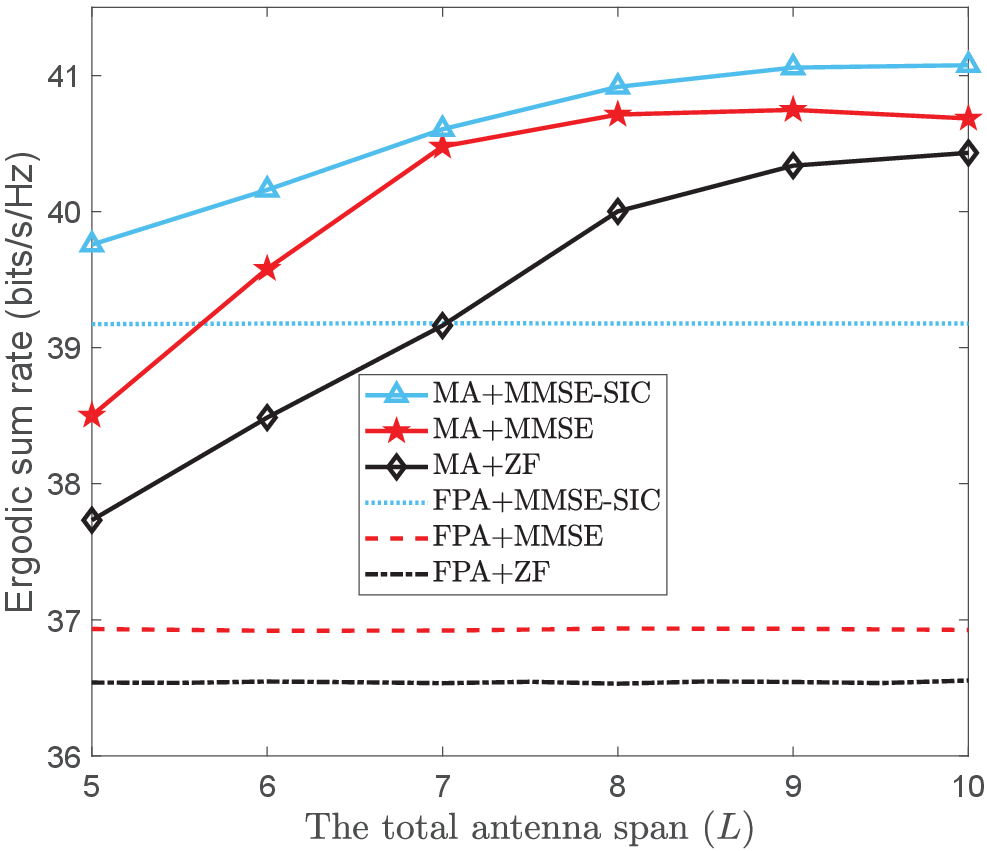}
\captionsetup{font=small}
\caption{Ergodic sum rate versus the total antenna span.} \label{fig:Fig1}
\vspace{-5pt}
\end{figure}

\section{Simulation Results}
In this section, we provide numerical results to validate the effectiveness of our proposed designs. Unless otherwise stated, the large-scale fading coefficients are set as $\left\{ {{\beta _m}} \right\}_{m = 1}^M = 1$, such that the channel power gains are normalized over the noise power and then we can set ${\sigma ^2} = 1$. The transmit power of each user is set as ${P_m} = 15$ dB, $m = 1,2,...,M$, and the Rician K-factors are $\left\{ {{K_m}} \right\}_{m = 1}^M = 10$. The number of users is $M = 5$, with $\left\{ {{\theta _m}} \right\}_{m = 1}^5 = \left\{ {0.0542, \ 0.8186, \ 0.9386, \ 0.2841, \ 0.1805} \right\}$. The minimum distance between any two adjacent MAs is set as ${d_{\min }} = 0.5\lambda $, with $\lambda  = 1$ for simplification. The proposed three receiving schemes armed with the MA technology are denoted as ``MA+ZF'' and ``MA+MMSE'' and ``MA+MMSE-SIC'', respectively. In addition, for the FPA array, the distance between any two adjacent antennas is fixed as $0.5\lambda $. The schemes which combine the FPA array with ZF, MMSE and MMSE-SIC are denoted as ``FPA+ZF'', ``FPA+MMSE'' and ``FPA+MMSE-SIC'', respectively.

Fig. 2 first illustrates the ergodic sum rate of different schemes versus the total antenna span $L$, where $N = 10$. From Fig. 2 we observe that: i) as $L$ increases in a small range, the ergodic sum rate of our proposed schemes will increase accordingly. This is because a larger $L$ provides more flexible space for antennas to move, such that the spatial DoF is significantly enhanced. However, when $L$ continues to increase, the performance will converge to a stationary value. This behavior implies that it is not necessary to infinitely expand the antenna span and only a finite $L$ is enough to achieve the performance bound; ii) as a comparison, the performance of the schemes based the FPA array is poor and not related to $L$. For instance, when $L = 8$, there exists about a 3.5 bits/s/Hz rate gap between ``MA+ZF'' and ``FPA+ZF'', indicating the great potential of the MA technology for enhancing the rate performance.

 \begin{figure}
 %\vspace{-10pt}
\centering
\includegraphics[width=6cm]{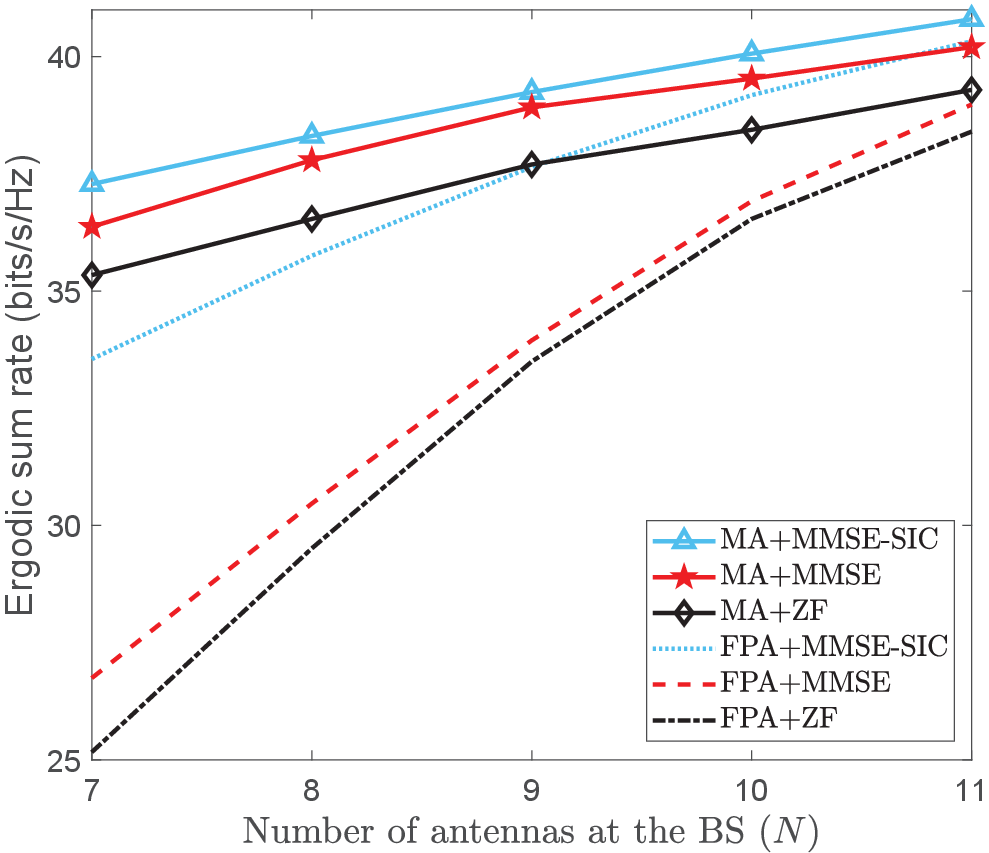}
\captionsetup{font=small}
\caption{Ergodic sum rate versus the number of antennas at the BS.} \label{fig:Fig1}
\vspace{-5pt}
\end{figure}

Fig. 3 further presents the rate performance versus the number of antennas ($N$) at the BS, where $L = 6$. From Fig. 3 we observe that as $N$ increases, i) on one hand, the BS can reap a higher reception diversity and then the performance of all schemes clearly becomes better; ii) on the other hand, the rate gap between the schemes armed with the MA technology and those based on the FPA array becomes smaller. This is because the total antenna span is fixed, and when $N$ is larger, there exists less space for each antenna to move and then the spatial DoF is reduced.

%\vspace{-5pt}
\section{Conclusion}
%\vspace{-5pt}
This letter studied the MA array-enabled multiuser uplink over general Rician fading channels. Considering practical implementations, a two-timescale design was proposed, where antenna positions were first optimized based on the statistical CSI and then the receiving beamforming (ZF, MMSE and MMSE-SIC) was designed with the instantaneous CSI. The formulated problems were highly non-convex and PGA algorithms were accordingly developed to solve them. Simulation results revealed the great potential of the MA technology for enhancing the ergodic sum rate.

\normalem
\bibliographystyle{IEEEtran}
\bibliography{IEEEabrv,mybib}

\end{document}